\documentclass[10pt, conference, letterpaper]{IEEEtran}
\usepackage{mathtools}
\usepackage[boxruled,vlined]{algorithm2e}
\usepackage{blindtext, graphicx}
\usepackage{url}
\usepackage{relsize}
\usepackage{cite}
\usepackage[breaklinks,colorlinks=true,citecolor=blue,pdftex]{hyperref}
\usepackage{bm}
\usepackage[long,nocomma]{optidef}
\usepackage{mdframed}
\usepackage{amsmath}
\usepackage{changepage}
\usepackage{subfigure}
\usepackage{anyfontsize}

\usepackage{array}

\IEEEoverridecommandlockouts

\begin{document}

\title{QuickCast: Fast and Efficient Inter-Datacenter Transfers using Forwarding Tree Cohorts\thanks{This is an extended version of a paper accepted for publication in IEEE INFOCOM 2018, Honolulu, HI, USA}}

\author{}

\author{\IEEEauthorblockN{Mohammad Noormohammadpour$^1$, Cauligi S. Raghavendra$^1$, Srikanth Kandula$^2$, Sriram Rao$^2$}
\IEEEauthorblockA{$^1$Ming Hsieh Department of Electrical Engineering, University of Southern California\\ $^2$Microsoft}}

\maketitle

\begin{abstract}
Large inter-datacenter transfers are crucial for cloud service efficiency and are increasingly used by organizations that have dedicated wide area networks between datacenters. A recent work uses multicast forwarding trees to reduce the bandwidth needs and improve completion times of point-to-multipoint transfers. Using a single forwarding tree per transfer, however, leads to poor performance because the slowest receiver dictates the completion time for all receivers. Using multiple forwarding trees per transfer alleviates this concern--the average receiver could finish early; however, if done naively, bandwidth usage would also increase and it is apriori unclear how best to partition receivers, how to construct the multiple trees and how to determine the rate and schedule of flows on these trees. This paper presents QuickCast, a first solution to these problems. Using simulations on real-world network topologies, we see that QuickCast can speed up the average receiver's completion time by as much as $10\times$ while only using $1.04\times$ more bandwidth; further, the completion time for all receivers also improves by as much as $1.6\times$ faster at high loads.
\end{abstract}



\section{Introduction}
Software Defined Networking (SDN) is increasingly adopted across Wide Area Networks (WANs) \cite{gartner-sdwan}. SDN allows careful monitoring, management and control of status and behavior of networks offering improved performance, agility and ease of management. Consequently, large cloud providers, such as Microsoft \cite{azure} and Google \cite{google}, have built dedicated large scale WAN networks that can be operated using SDN which we refer to as SD-WAN. These networks connect dozens of datacenters for increased reliability and availability as well as improved utilization of network bandwidth reducing communication costs \cite{swan, b4}.

Employing geographically distributed datacenters has many benefits in supporting users and applications. Replicating objects across multiple datacenters improves user-access latency, availability and fault tolerance. For example, Content Delivery Networks (CDNs) replicate objects (e.g. multimedia files) across many cache locations, search engines distribute large index updates across many locations regularly, and VMs are replicated across multiple locations for scale out of applications. In this context, \textit{Point to Multipoint (P2MP)} transfers (also known as One-to-Many transfers) are necessary. A P2MP transfer is a special case of multicasting with a single sender and a fixed set of receivers that are known apriori. These properties together provide an opportunity for network optimizations, such as sizable reductions in bandwidth usage and faster completion times by using carefully selected forwarding trees. 

We review several approaches for performing P2MP transfers. One can perform P2MP transfers as many independent point-to-point transfers \cite{tempus, owan, swan, b4} which can lead to wasted bandwidth and increased completion times. Internet multicasting approaches \cite{ip_multicast} build multicast trees gradually as new receivers join the multicast sessions, and do not consider the distribution of load across network links while connecting new receivers to the tree. This can lead to far from optimal multicast trees which are larger than necessary and poor load balancing. Application layer multicasting, such as \cite{nice}, focuses on use of overlay networks for building virtual multicast trees. This may lead to poor performance due to limited visibility into network link level status and lack of control over how traffic is directed in the network. Peer-to-peer file distribution techniques \cite{slurpie, bittorrent} try to maximize throughput per receiver locally and greedily which can be far from a globally optimal solution. Centralized multicast tree selection approaches have been proposed \cite{avalanche, datacast} that operate on regular and structured topologies of networks inside datacenters which cannot be directly applied to inter-DC networks. Other related research either does not consider elasticity of inter-DC transfers which allows them to change their transmission rate according to available bandwidth \cite{raera, sdn_multicast} or inter-play among may inter-DC transfers for global network-wide optimization \cite{MPMC_2013, MPMC_2016}.

We recently presented a solution called DCCast \cite{dccast} that reduces tail completion times for P2MP transfers. DCCast employs a central traffic engineering server with a global view of network status, topology and resources to select forwarding trees over which traffic flows from senders to all receivers. This approach combines good visibility into and control over networks offered by SDN with bandwidth savings achieved by reducing the number of links used to make data transfers. OpenFlow \cite{openflow}, a dominant SDN protocol, has supported forwarding to multiple outgoing ports since $\mathrm{v}1.1$ \cite{openflow-1.1.0}; for example, by using Group Tables and the \texttt{OFPGT\_ALL} flag. Several recent SDN switches support this feature  \cite{of-gt-1, of-gt-2, of-gt-3, of-gt-4}.

In this paper, we propose a new rate-allocation and tree selection technique called \textbf{QuickCast} with the aim of \textit{minimizing average completion times} of inter-DC transfers. QuickCast reduces completion times by replacing a large forwarding tree with multiple smaller trees each connected to a subset of receivers which we refer to as a cohort of forwarding trees. Next, QuickCast applies the Fair Sharing scheduling policy which we show through simulations and examples minimizes contention for available bandwidth across P2MP transfers compared to other scheduling policies.

\begin{figure}
\centering
\includegraphics[width=\columnwidth]{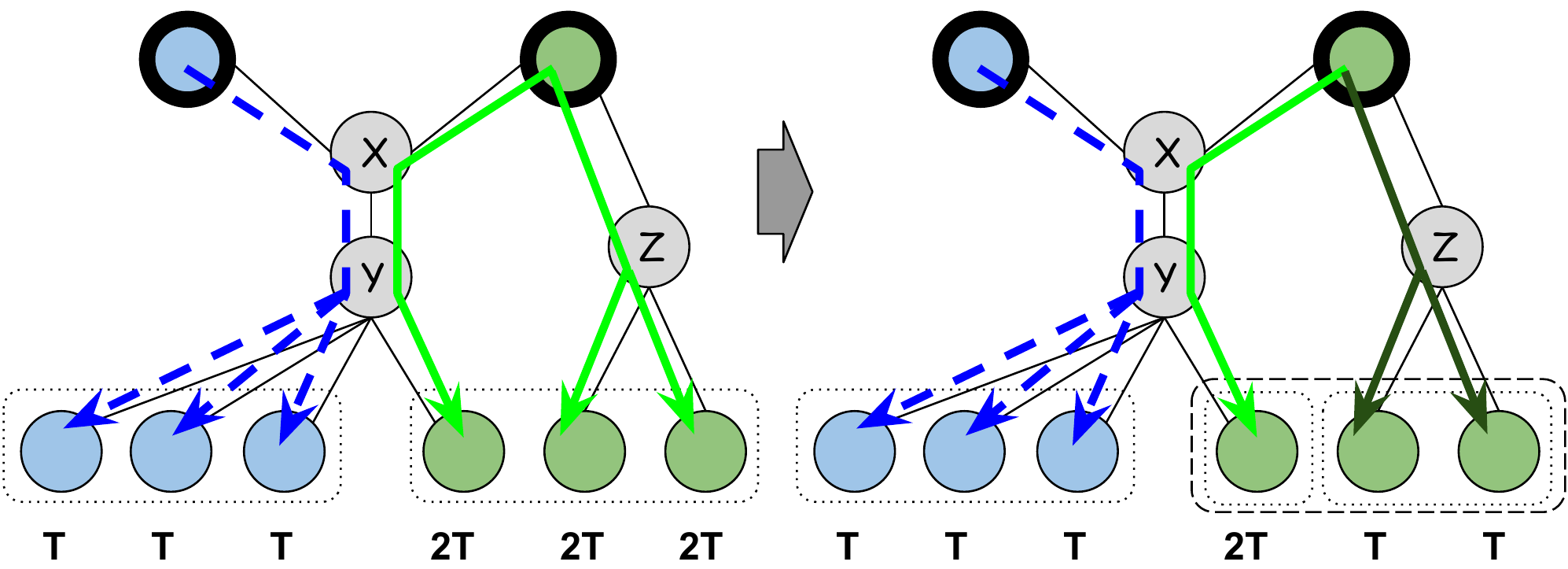}
\caption{Partitioning receivers into subsets can improve mean completion times (this figure assumes FCFS rate-allocation policy, dashed blue transfer arrives slightly earlier, all links have equal capacity of $1$)}
\label{fig_example}
\end{figure}

Despite offering bandwidth savings and reduced tail completion times, DCCast can suffer from significant increase in completion times when P2MP transfers have many receivers. As the number of receivers increases, forwarding trees grow large creating many overlapping edges across transfers, increasing contention and completion times. Since the edge with minimal bandwidth determines the overall throughput of a forwarding tree, we refer to this as \textbf{``Weakest Link"} problem. We demonstrate weakest link problem using a simple example. Figure \ref{fig_example} shows a scenario where two senders (top two nodes) are transmitting over forwarding trees and they share a link ($x \rightarrow y$). According to DCCast which uses the First Come First Serve (FCFS) policy, the dashed blue transfer (on left) which arrived just before the green (on right) is scheduled first delaying the beginning of green transfer until time $T$. As a result, all green receivers finish at time $2T$. By using multiple trees, in this case two for the green sender, each tree can be scheduled independently, and thereby reducing completion time of the two receivers on the right from $2T$ to $T$. That is, we create a new tree that does not have the \textit{weakest} link, which in this scenario is $x \rightarrow y$ for both blue and green transfers.

\begin{figure}
\centering
\includegraphics[width=0.95\columnwidth]{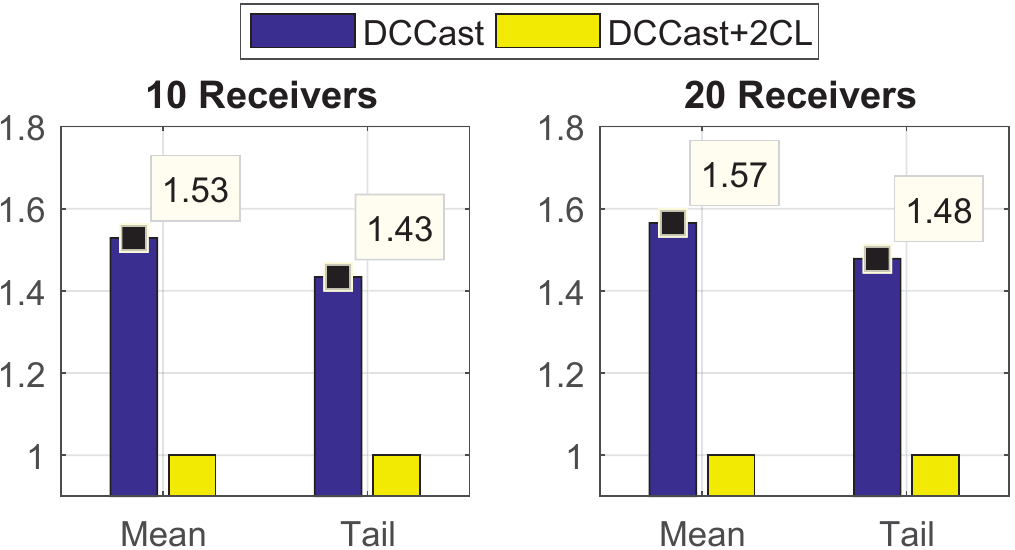}
\caption{Comparison of completion times (mean and tail) of DCCast vs. DCCast+2CL (two-clustering according to proximity) for P2MP transfers with $10$ and $20$ receivers per transfer (Normalized)}
\label{fig_improve}
\end{figure}

To replace a large tree with a cohort of trees, we propose partitioning receiver sets of P2MP transfers into multiple subsets and using a separate forwarding tree per subset. This approach can significantly reduce completion times of P2MP transfers. We performed an experiment with DCCast over random topologies with $50$ nodes to determine if there is benefit in partitioning all P2MP transfers. We simply grouped receivers into two subsets according to proximity, i.e., shortest path hop count between receiver pairs, and attached each partition with an independent forwarding tree (DCCast+2CL). As shown in Figure \ref{fig_improve}, this reduced completion times by $50\%$ while increasing bandwidth usage by $6\%$ (not shown).

To further reduce completion times and decrease additional bandwidth usage, partitioning should be only applied to P2MP transfers that benefit from it. For example, using more than one tree for the dashed blue transfer (on left) in Figure \ref{fig_example} will increase contention and will hurt completion times. By carefully selecting P2MP transfers that benefit from partitioning and using an independent forwarding tree per partition, we can considerably improve average completion times with small extra bandwidth usage. We refer to this approach as \textbf{selective partitioning}. We limit the number of partitions per transfer to two to minimize bandwidth overhead of additional edges, minimize contention due to overlapping trees and limit the number of network forwarding entries.

In addition to using multiple forwarding trees, we investigate various scheduling policies namely FCFS used in DCCast, Shortest Remaining Processing Time (SRPT) and Fair Sharing based on Max-Min Fairness (MMF) \cite{max-min-fairness}. Although SRPT is optimal for minimizing mean completion times while scheduling traffic on a single link, we find that MMF beats SRPT by a large factor as forwarding trees grow (which causes many trees to share every link) and as offered load increases (which increases contention for using available bandwidth).

Using a cohort of forwarding trees per P2MP transfer can also increase reliability in two ways. First, by mitigating the effect of weak links, a number of receivers will complete reception earlier which reduces the probability of data loss due to failures. Second, in case of a link failure, using more than one tree reduces the probability of all receivers being affected. In case subsequent trees are constructed in a link disjoint manner, no single link failure can affect all receivers. In summary, we make the following contributions:

\begin{enumerate}
    \item We formulate a general optimization problem for minimizing mean completion times while scheduling P2MP transfers over a general network.
    \item We propose QuickCast, which mitigates the ``Weakest Link" problem by partitioning receivers into multiple subsets and attaching each partition with an independent forwarding tree. Using multiple forwarding trees can improve average completion time and reliability with only a little additional bandwidth usage. We show that partitioning should only be applied to transfers that benefit from it, i.e., partitioning some transfers leads to increased contention and worse completion times.
    \item We explore well-known scheduling policies (FCFS, SRPT and Fair Sharing) for central rate-allocation over forwarding trees and find MMF to be the most effective in maximizing overall bandwidth utilization and reducing completion times.
    \item We perform extensive simulations using both synthetic and real inter-datacenter traffic patterns over real WAN topologies. We find that performance gain of QuickCast depends on the offered load and show that under heavy loads, QuickCast can improve mean times by as much as $10\times$ and tail times by as much as $1.57\times$ while imposing very small increase in typical bandwidth usage (only $4\%$) compared to DCCast.
\end{enumerate}

The rest of this paper is organized as follows. In Section \ref{problem}, we state the P2MP scheduling problem, explain the constraints and variables used in the paper and formulate the problem as an optimization scenario. In Section \ref{quickcast}, we present QuickCast and the two procedures it is based on. In Section \ref{eval}, we perform abstract simulations to evaluate QuickCast and in Section \ref{discussion} we discuss practical considerations and issues. At the end, in Section \ref{related_works}, we provide a comprehensive overview of related research efforts.

\section{Assumptions and Problem Statement} \label{problem}
Similar to prior work on inter-datacenter networks \cite{b4, swan, tempus, amoeba, dcroute, dccast, owan}, we assume a SD-WAN managed by a Traffic Engineering (TE) server which receives transfer requests from end-points, performs rate-allocations and manages the forwarding plane. Transfers arrive at the TE server in an online fashion and are serviced as they arrive. Requests are specified with four parameters of arrival time, source, set of receivers and size (volume in bytes). End-points apply rate-limiting to minimize congestion. We consider a slotted timeline to allow for flexible rate-allocation while limiting number of rate changes to allow time to converge to specified rates and minimize rate-allocation overhead \cite{tempus, amoeba}. We focus on long running transfers that deliver large objects to many datacenters such as applications in \cite{b4}. For such transfers, small delays are usually acceptable, including overhead of centralized scheduling and network configurations. To reduce configuration overhead (e.g. latency and control plane failures \cite{ffc}), we assume that a forwarding tree is not changed once configured on the forwarding plane.

To reduce complexity we assume that end-points can accurately rate-limit data flows and that they quickly converge to required rates; that there are no packet losses due to congestion, corruption or errors; and that scheduling is done for a specific class of traffic meaning all requests have the same priority. In Section \ref{discussion}, we will discuss ways to deal with cases when some of these assumptions do not hold.

\begin{table}
\caption{Definition of variables} \label{table_var}
\begin{center}
\begin{tabular}{ |p{1.7cm}|p{6.2cm}| }
    \hline
    \textbf{Variable} & \textbf{Definition} \\
    \hline
    \hline
    $t$ and $t_{now}$ & Some timeslot and current timeslot \\
    \hline
    $N$ & Total number of receivers per transfer \\
    \hline
    $e$ & A directed edge \\
    \hline
    $C_e$ & Capacity of $e$ in bytes per second \\
    \hline
    $(x, y)$ & A directed edge from $x$ to $y$ \\
    \hline
    $G$ & A directed inter-datacenter network graph \\
    \hline
    $T$ & Some directed tree connecting a sender to its receivers \\
    \hline
    $\pmb{\mathrm{V_G}}$ and $\pmb{\mathrm{V_T}}$ & Set$\langle\rangle$ of vertices of $G$ and $T$ \\
    \hline
    $\pmb{\mathrm{E_G}}$ and $\pmb{\mathrm{E_T}}$ & Set$\langle\rangle$ of edges of $G$ and $T$ \\
    \hline
    $B_e$ & Current available bandwidth on edge $e$ \\
    \hline
    $B_T$ & Current available bandwidth over tree $T$ \\
    \hline
    $\delta$ & Width of a timeslot in seconds \\
    \hline
    $R_i$ & A transfer request where $i \in \pmb{\mathrm{I}}=\{1 \dots I\}$ \\
    \hline
    $O_{i}$ & Data object associated with $R_i$ \\
    \hline
    $S_{R_i}$ & Source datacenter of ${R_i}$ \\
    \hline
    $A_{R_i}$ & Arrival time of ${R_i}$ \\
    \hline
    $\mathcal{V}_{R_i}$ & Original volume of ${R_i}$ in bytes \\
    \hline
    $\mathcal{V}_{R_i}^r$ & Residual volume of ${R_i}$,  ($\mathcal{V}_{R_i}^r=\mathcal{V}_{R_i}$ at $t=A_{R_i}$) \\
    \hline
    $\pmb{\mathrm{D}}_{R_i}$ & Set$\langle\rangle$ of destinations of ${R_i}$ \\
    \hline
    $n$ & Maximum subsets (partitions) allowed per receiver set \\ 
    \hline
    $\pmb{\mathrm{P}}_{i}^{j}$ & Set$\langle\rangle$ of receivers of $R_i$ in partition $j \in \{1,\dots,n\}$ \\
    \hline
    $q_{i}$ & Total bytes used to deliver $O_{i}$ to $\pmb{\mathrm{D}}_{R_{i}}$ \\
    \hline
    $T_{i}^{j}$ & Forwarding tree for partition $j \in \{1,\dots,n\}$ of $R_i$ \\
    \hline
    $L_e$ & $e$'s total outstanding load, i.e., $L_e = \sum_{\substack{i,j\\e \in T_i^j}} {\cal V}_{R_i}^r$ \vspace{0.2em} \\
    \hline
    $f_{i}^{j}(t)$ & Transmission rate of $R_{i}$ on $T_{i}^{j}$ at timeslot $t$ \\
    \hline
    $\gamma_{i}^{j}(t)$ & Whether $R_{i}$ is transmitted over $T_{i}^{j}$ at timeslot $t$ \\
    \hline
    $\theta_{i,e}^{j}$ & Whether edge $e \in \pmb{\mathrm{E_G}}$ is on $T_{i}^{j}$ \\
    \hline
    $\nu_{i,j,v}$ & Whether $v \in \pmb{\mathrm{D}}_{R_{i}}$ is in $\pmb{\mathrm{P}}_{i}^{j}$ \\
    \hline
    $\pmb{\mathrm{M}}^{j}_{i}$ & $\{ \pmb{\mathrm{\nabla}} ~\vert~ \pmb{\mathrm{\nabla}} \subset \pmb{\mathrm{V_G}},~ \pmb{\mathrm{\nabla}} \cap \{\pmb{\mathrm{P}}_{i}^{j} \cup S_{R_{i}}\} \neq \emptyset, (\pmb{\mathrm{V_G}}-\pmb{\mathrm{\nabla}}) \cap \{\pmb{\mathrm{P}}_{i}^{j} \cup S_{R_{i}}\} \neq \emptyset \}$ \\
    \hline
    $E(\pmb{\mathrm{\nabla}})$ & $\{e=(x,y) ~\vert~ x \in \pmb{\mathrm{\nabla}}, y \in (\pmb{\mathrm{V_G}}-\pmb{\mathrm{\nabla}})\}$ \\
    \hline
\end{tabular}
\end{center}
\end{table}

\begin{figure}
\begin{mdframed}[userdefinedwidth=0em,
                align=center,
                skipabove=0,
                skipbelow=0]
{\footnotesize
\begin{mini*}[3]
    {}{\sum_{i \in \pmb{\mathrm{I}}} \Big(\sum_{j \in \{1,\dots,n\}} \lvert \pmb{\mathrm{P}}_{i}^{j} \rvert ~\big(\sum_{t > A_{R_{I+1}}} (t-A_{R_{I+1}}) \gamma_{i}^{j}(t) }{}{}
    \breakObjective{\prod_{t^{\prime} > t} (1-\gamma_{i}^{j}(t^{\prime}))~\big)\Big)}
    \breakObjective{+ \Big(\frac{\sum_{i \in \pmb{\mathrm{I}}} q_{i}}{n ~ \lvert \pmb{\mathrm{E_G}}\rvert  ~  \sum_{i \in \pmb{\mathrm{I}}} \mathcal{ V}_{R_{i}}}\Big)}
    {}{}
    \addConstraint{}{\text{Calculate total bandwidth usage:}}{}
    \addConstraint{(1)\ \ }{q_{i}=\sum_{e \in \pmb{\mathrm{E_G}}} (\sum_{j \in \{1,\dots,n\}} \theta_{i,e}^{j}) \mathcal{V}_{R_{i}}^r \qquad\qquad\qquad}{\forall i}
    \addConstraint{}{\text{Demand satisfaction constraints:}}{}
    \addConstraint{(2)\ \ }{\sum_{t} \gamma_{i}^{j}(t) f_{i}^{j}(t)=\frac{\mathcal{ V}_{R_{i}}^r}{\delta}}{  \forall i,j}
    \addConstraint{}{\text{Capacity constraints:}}{}
    \addConstraint{(3)\ \ }{\sum_{i \in \pmb{\mathrm{I}}} \sum_{j \in \{1,\dots,n\}} \theta_{i,e}^{j} f_{i}^{j}(t) \le C_e}{ \forall j,t,e}
    \addConstraint{}{\text{Steiner tree constraints \cite{steiner_mip}:}}{}
    \addConstraint{(4)\ \ }{\sum_{e \in E(\pmb{\mathrm{\nabla}})} \theta_{i,e}^{j} \ge 1}{\forall i,j, \nabla \in \pmb{\mathrm{M}}^{j}_{i}}
    %
    %
    \addConstraint{}{\text{Basic range constraints:}}{}
    \addConstraint{(5)\ \ }{\gamma_{i}^{j}(t) = 0,~ f_{i}^{j}(t) = 0}{\forall i,j,t < A_{R_{I+1}}}
    \addConstraint{(6)\ \ }{f_{i}^{j}(t) \ge 0}{\forall i,j,t}
    \addConstraint{(7)\ \ }{\theta_{i,e}^{j} \in \{0,1\}}{\forall i,j,e}
    \addConstraint{(8)\ \ }{\gamma_{i}^{j}(t) \in \{0,1\}}{\forall i,j,t}
    \addConstraint{(9)\ \ }{\nu_{i,j,v} \in \{0,1\}}{\forall i,j,v \in \pmb{\mathrm{D}}_{R_{i}}}
    \addConstraint{(10)\ \ }{\gamma_{i}^{j}(t) = 0}{\forall i,j,t < A_{R_{i}}}
\end{mini*}
}
\end{mdframed}
\caption{Online optimization model, variables defined in Table \ref{table_var}} \label{eq_opt_model}
\end{figure}

\subsection{Problem Formulation} \label{optimization}
Earlier we showed that partitioning of receiver sets can improve completion times via decreasing network contention. However, to further improve completion times, partitioning should be done according to transfer properties, topology and network status. Optimal partitioning for minimization of completion times is an open problem and requires finding solution to a complex joint optimization model that takes into account forwarding tree selection and rate-allocation.

We first point out the most basic constraint out of using forwarding trees. Table \ref{table_var} provides the list of variables we will use. For any tree $T$, packets flow from the source to receivers with same rate $r_T$ that satisfies:

\vspace{-0.8em}
\begin{align}
    r_T &\le B_T = \min_{e \in \pmb{\mathrm{E_T}}}(B_e) \label{eq_rate_1}
\end{align}

We formulate the problem as an online optimization scenario. Assuming a set of requests $R_i, ~i \in \{1 \dots I\}$ already in the system, upon arrival of new request $R_{I+1}$, an optimization problem needs to be formulated and solved to find rate-allocations, partitions and forwarding trees. Figure \ref{eq_opt_model} shows the overall optimization problem. This model considers up to $n \ge 1$ partitions per transfer. Demands of existing requests are updated to their residuals upon arrival of new requests.

The objective is formulated in a hierarchical fashion giving higher priority to minimizing mean completion times and then reducing bandwidth usage. The purpose of $\gamma_{i}^{j}(t)$ indicator variables is to calculate the mean times: the latest timeslot over which we have $f_{i}^{j}(t) > 0$ determines the completion time of partition $j$ of request $i$. These completion times are then multiplied by partition size to create the total sum of completion times per receiver. Constraint $4$ ascertains that there is a connected subgraph across sender and receivers per partition per request which is similar to constraints used to find minimal edge Steiner trees \cite{steiner_mip}. Since our objective is an increasing function of bandwidth, these constraints eventually lead to minimal trees connecting any sender to its receivers while not increasing mean completion times (the second part of objective that minimizes bandwidth is necessary to ensure no stray edges).

\subsection{Challenges}
We focus on two objectives of first minimizing completion times of data transfers and minimizing total bandwidth usage. This is a complex optimization problem for a variety of reasons. First, breaking a receiver set to several partitions leads to exponential number of possibilities. Moreover, the optimization version of Steiner tree problem which aims to find minimal edge or minimal weight trees is a hard problem \cite{steiner_mip}. Completion times of transfers then depend on how partitions are formed and which trees are used to connect partitions to senders. In addition, the scenario is naturally an online problem which means even if we were able to compute an optimal solution for a given set of transfers in a short amount of time, we still would not be able to compute a solution that is optimal over longer periods of time due to incomplete knowledge of future arrivals.

\section{QuickCast} \label{quickcast}
We present our heuristic approach in Algorithm \ref{algo_quick} called QuickCast with the objective of reducing mean completion times of elastic P2MP inter-DC transfers. We first review concepts behind design of this heuristic, namely rate-allocation, partitioning, forwarding tree selection and selective partitioning. Next, we discuss how Algorithm \ref{algo_quick} realizes these concepts using two procedures one executed upon arrival of new transfers and the other per timeslot.

\subsection{Rate-allocation}
To compute rates per timeslot, we explore well-known scheduling policies: FCFS, SRPT and Fair Sharing. Although simple, FCFS can lead to increased mean times if large transfers block multiple edges by fully utilizing them. SRPT is known to offer optimal mean times over a single link but may lead to starvation of large transfers. QuickCast uses Fair Sharing based on MMF policy.

\begin{figure}
\centering
\includegraphics[width=\columnwidth]{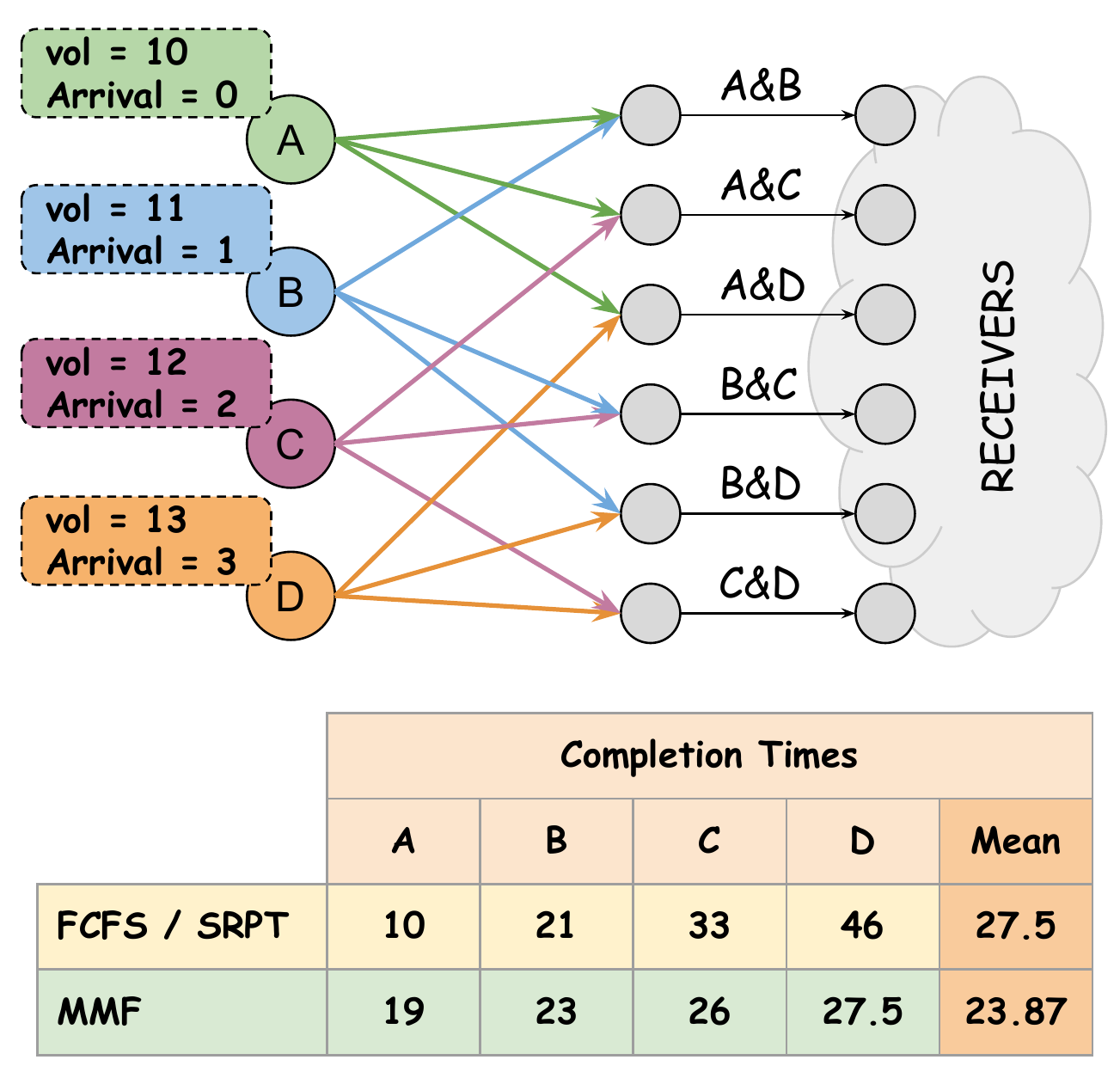}
\caption{Fair Sharing can offer better mean times compared to both SRPT and FCFS while scheduling over forwarding trees, all links have capacity of $1$}
\label{fig_fs_srpt}
\end{figure}

To understand effect of different scheduling policies, let us consider the following example. Figure \ref{fig_fs_srpt} shows a scenario where multiple senders have initiated trees with multiple branches and they share links along the way to their receivers. SRPT gives a higher priority to the top transfer with size $10$ and then to the next smallest transfer and so on. When the first transfer is being sent, all other transfers are blocked due to shared links. This occurs again when the next transfer begins. Scheduling according to FCFS leads to same result. In this example, mean completion times for both FCFS and SRPT is about $1.16\times$ larger than MMF. In Section \ref{eval}, we perform simulation experiments that confirm the outcome in this example. We find that as trees grow larger and under high utilization, the benefit of using MMF over FCFS or SRPT becomes more significant due to increased contention. We also realize that tail times grow much faster for SRPT compared to both MMF and FCFS (since it also suffers from the starvation problem) while scheduling over forwarding trees with many receivers, and that increases SRPT's mean completion times.

\subsection{Partitioning}
There are two configuration parameters for grouping receivers into multiple partitions: \textit{partitioning criteria} and \textit{number of partitions}. In general, partitioning may lead to higher bandwidth usage. However, it may be the case that a small increase in bandwidth usage can considerably improve completion times. Efficient and effective partitioning to minimize completion times and bandwidth usage is a complex open problem. We discuss our approach in the following.

\textbf{Partitioning Criteria:} We focus on minimizing extra bandwidth usage while breaking large forwarding trees via partitioning. Generally, one could select partitions according to current network conditions, such as distribution of load across edges. However, we notice that network conditions are continuously changing as current transfers finish and new transfers arrive. Minimizing bandwidth on the other hand appears as a globally desirable objective for partitioning and was hence chosen. To find partitions, QuickCast groups receivers according to proximity until we are left with desired number of groups each forming a partition. Distance between two receivers is computed as the number of hops on the shortest path between them. With this approach, a partition requires minimal number of edges to connect the nodes within. Reassigning any receiver to other partitions will increase the number of edges and thus consumed bandwidth. 

\textbf{Number of Partitions:} The right number of partitions per transfer depends on factors such as topology, number of receivers, forwarding trees of other transfers, source and destinations of a transfer, and overall network load. In the extreme case of $N$ partitions, a P2MP transfer is broken into $N$ unicast transfers which significantly increases bandwidth usage. Partitioning is most effective if forwarding trees assigned to partitions do not increase overall contention for network bandwidth, i.e., the number of overlapping edges across new trees is minimal. Therefore, increasing number of partitions to more than connectivity degree of datacenters may offer minor gains or even loss of performance (e.g., in case of Google B4 \cite{b4}, the minimum and maximum connectivity degrees are $2$ and $4$, respectively). From the practical aspect, number of partitions and hence forwarding trees determines the number of Group Table rules that need to be setup in network switches. Therefore, we focus on partitioning receivers into up to $2$ groups each assigned an independent forwarding tree. Exploration of effects of more partitions is left for future work.

\subsection{Forwarding Tree Selection}
After computing two partitions, QuickCast assigns an independent forwarding tree per partition using tree selection approach presented in \cite{dccast} which was shown to provide high bandwidth savings and improved completion times. It operates by giving a weight of $W_e = (L_{e} + {\cal V}_R)$ (see Table \ref{table_var} for definition of variables) and then selecting the minimum weight forwarding tree. This technique allows load balancing of P2MP data transfers over existing trees according to total bandwidth scheduled on the edges. It also takes into account transfer volumes while selecting trees. Particularly, larger transfers are most likely assigned to smaller trees to minimize bandwidth usage while smaller transfers are assigned to least loaded trees (regardless of tree size). This approach becomes more effective when a small number of transfers are orders of magnitude larger than median \cite{social_inside}, as number of larger forwarding trees is usually significantly larger than smaller trees on any graph.

\subsection{Selective Partitioning}
Partitioning is beneficial only if it decreases or minimally increases bandwidth usage and contention over resources which necessitates selectively partitioning the receivers. When we chose the two partitions by grouping receivers and after selecting a forwarding tree for every group, QuickCast calculates the total weight of each forwarding tree by summing up weights of their edges. We then compare sum of these two weights with no partitioning case where a single tree was used. If the total weight of two smaller trees is less than some \textit{partitioning factor} (shown as $p_f$) of the single tree case, we accept to use two trees. If $p_f$ is close to $1.0$, partitioning occurs only if it incurs minimal extra weight, i.e., $(p_f - 1)$ times weight of the single forwarding tree that would have been chosen if we applied no partitioning. With this approach, we most likely avoid selection of subsequent trees that are either much larger or much more loaded than the initial tree in no partitioning case. Generally, an effective $p_f$ is a function of traffic distribution and topology. According to our experiments with several traffic distributions and topologies, choosing it in the range of $1.0 \le p_f \le 1.1$ offers the best completion times and minimal bandwidth usage.

\subsection{QuickCast Algorithm}
A TE server is responsible for managing elastic transfers over inter-DC network. Each partition of a transfer is managed independently of other partitions. We refer to a transfer partition as \textbf{active} if it has not been completed yet. TE server keeps a list of active transfer partitions and tracks them at every timeslot. A TE server running QuickCast algorithm uses two procedures as shown in Algorithm \ref{algo_quick}.

\textbf{Submit($\mathrm{R,n,p_f}$):} This procedure is executed upon arrival of a new P2MP transfer $R$. It performs partitioning and forwarding tree selection for the new transfer given its volume, source and destinations. We consider the general case where we may have up to $n$ partitions per transfer. First, we compute edge weights based on current traffic matrix and volume of new transfer. We then build the agglomerative hierarchy of receivers using average linkage and considering proximity as clustering metric. Agglomerative clustering is a bottom up approach where at every level the two closest clusters are merged forming one cluster. The distance of any two clusters is computed using average linkage which is the average over pairwise distances of nodes in the two clusters. The distance between every pair of receivers is the number of edges on the shortest path from one to the other. It should be noted that although our networks are directed, all edges are considered to be bidirectional and so the distance in either direction between any two nodes should be the same. When the hierarchy is ready, we start from the level where there are $n$ clusters (or at the bottom if total number of receivers is less than or equal to $n$) and compute the total weight of $n$ forwarding trees (minimum weight Steiner trees) to these clusters. We move forward with this partitioning if the total weight is less than $p_f$ times weight of the forwarding tree that would have been selected if we grouped all receivers into one partition. Otherwise, the same process is repeated while moving up one level in the clustering hierarchy (one less cluster). If we accept a partitioning, this procedure first assigns a forwarding tree to every partition while continuously updating edge weights. It then returns the partitions and their forwarding trees.

\textbf{DispatchRates():} This procedure is executed at the beginning of every timeslot. It calculates rates per active transfer partition and according to MMF rate-allocation policy. New transfers arriving somewhere within a timeslot are allocated rates starting next timeslot. To calculate residual demands needed for rate calculations, senders report back the actual volume of data delivered during past timeslot per partition. This allows QuickCast to cope with inaccurate rate-limiting and packet losses which may prevent a transfer from fully utilizing its allotted share of bandwidth.

\SetAlgoVlined
\SetInd{1.2em}{0.5em}
\begin{algorithm}
\caption{QuickCast} \label{algo_quick}
\fontsize{9.5}{11.4}\selectfont
\SetKw{KwBy}{by}
\SetKwProg{Submit}{Submit}{}{}

\Submit{$\mathrm{(R,n,p_f)}$}{
    \vspace{0.4em}
    \KwIn{$\mathrm{R}({\cal V}_R,S_R,\pmb{\mathrm{D}}_R)$, $n$ ($=2$ in this paper), $p_f$, $G$, $L_{e}$ for $\forall e \in \pmb{\mathrm{E_G}}$ (Variables defined in Table \ref{table_var})}
    
    \vspace{0.4em}
    \KwOut{Pairs of (Partition, Forwarding Tree)}
    
    \vspace{0.4em}
    $\forall \alpha,\beta \in \pmb{\mathrm{D}}_R,~\alpha \neq \beta$, $\mathrm{DIST}_{\alpha,\beta} \gets$ number of edges on the shortest path from $\alpha$ to $\beta$\;
    
    \vspace{0.4em}
    To every edge $e \in \pmb{\mathrm{E_G}}$, assign weight $W_e = (L_{e} + {\cal V}_R)$\;
    
    \vspace{0.4em}
    Find the minimum weight Steiner tree $T_R$ that connects $S_R \cup \pmb{\mathrm{D}}_R$ and its total weight $W_{T_{R}}$\;
    
    \vspace{0.4em}
    \For{$k = n$ \KwTo $k = 2$ \KwBy $-1$}{
        Agglomeratively cluster $\pmb{\mathrm{D}}_R$ using average linkage and distance metric $\mathrm{DIST}$ calculated in previous line until only $k$ clusters left forming $\pmb{\mathrm{P}}_R^{i},~i \in \{1,\dots,k\}$\;
        
        \vspace{0.4em}
        \ForEach{$i \in \{1,\dots,k\}$}{
            \vspace{0.4em}
            Find $W_{T_{\pmb{\mathrm{P}}_R^{i}}}$, weight of minimum weight Steiner tree that connects $S_R \cup \pmb{\mathrm{P}}_R^{i}$\;
        }
        
        \vspace{0.4em}
        \If{$~\sum_{i \in \{1,\dots,k\}} W_{T_{\pmb{\mathrm{P}}_R^{i}}} \le p_f \times W_{T_{R}}~$}{
            \vspace{0.4em}
            \ForEach{$i \in \{1,\dots,k\}$}{
                \vspace{0.4em}
                Find the minimum weight Steiner tree $T_{\pmb{\mathrm{P}}_R^{i}}$ that connects $S_R \cup \pmb{\mathrm{P}}_R^{i}$\;
                \vspace{0.4em}
                $L_e \gets L_e + {\cal V}_R,~\forall e \in T_{\pmb{\mathrm{P}}_R^{i}}$\;
                \vspace{0.4em}
                Update $W_e = (L_{e} + {\cal V}_R)$ for all $e \in \pmb{\mathrm{E_G}}$\;
            }
            
            \vspace{0.4em}
            \Return{$\pmb{\mathrm{P}}_R^{i}$ as well as $T_{\pmb{\mathrm{P}}_R^{i}},~\forall i \in \{1,\dots,k\}$}\;
        }
    }
    
    $L_e \gets L_e + {\cal V}_R,~\forall e \in T_{R}$\;
    
    \vspace{0.4em}
    \Return{$\pmb{\mathrm{D}}_R$ and $T_{R}$}\;
}

\vspace{1.5em}
\SetKwProg{DispatchRates}{DispatchRates}{}{}

\DispatchRates{$\mathrm{()}$}{
    \vspace{0.4em}
    \KwIn{Set of active request partitions $\pmb{\mathrm{P}}$, their current residual demands and forwarding trees $\mathcal{V}_{P}^r$ and $T_{P},~\forall P \in \pmb{\mathrm{P}}$, and timeslot width $\delta$}
    
    \vspace{0.4em}
    \KwOut{Rate per active request per partition for next timeslot}
    
    \vspace{0.4em}
    $\mathrm{COUNT}_e \gets$ number of forwarding trees $T_{P},~\forall P \in \pmb{\mathrm{P}}$ sharing edge $e,~\forall e \in \pmb{\mathrm{E_G}}$\;
    
    \vspace{0.4em}
    $\pmb{\mathrm{P}}^\prime \gets \pmb{\mathrm{P}}$ and $\mathrm{CAP}_e \gets 1,~\forall e \in \pmb{\mathrm{E_G}}$\;
    
    \vspace{0.4em}
    \While{$\lvert \pmb{\mathrm{P}}^\prime \rvert > 0$}{
        \ForEach{$P \in \pmb{\mathrm{P}}^\prime$}{
            $\mathrm{SHARE}_{P} \gets \min_{e \in T_{P}}(\frac{\mathrm{CAP}_e}{\mathrm{COUNT}_e})$\;
        }
        
        \vspace{0.4em}
        $P^\prime \gets $ a partition $P$ with minimum $\mathrm{SHARE}_P$ value\;
        
        \vspace{0.4em}
        $\mathrm{RATE}_{P^\prime} \gets \min(\mathrm{SHARE}_{P^\prime}, \frac{\mathcal{V}_{P^\prime}^r}{\delta})$\;
        
        \vspace{0.4em}
        $\pmb{\mathrm{P}}^\prime \gets \pmb{\mathrm{P}}^\prime - \{P^\prime\}$\;
        
        \vspace{0.4em}
        $\mathrm{COUNT}_e \gets \mathrm{COUNT}_e - 1,~\forall e \in T_{P^\prime}$\;
        
        \vspace{0.4em}
        $\mathrm{CAP}_e \gets \mathrm{CAP}_e - \mathrm{RATE}_{P^\prime},~\forall e \in T_{P^\prime}$\;
    }
    
    \Return{$\mathrm{RATE}_{P},~\forall P \in \pmb{\mathrm{P}}$}
}

\end{algorithm}

\section{Evaluations} \label{eval}
We considered various topologies and transfer size distributions as in Tables \ref{table_topology} and \ref{table_traffic}. For simplicity, we considered a uniform capacity of $1.0$ for all edges, accurate rate-limiting at end-points, no dropped packets due to congestion or corruption, and no link failures. Transfer arrival followed a Poisson distribution with rate of $\lambda$. For all simulations, we considered a partitioning factor of $p_f = 1.1$ and timeslot length of $\delta = 1.0$. Unless otherwise stated, we assumed a fixed $\lambda=1.0$. Also, for all traffic distributions, we considered an average demand equal to volume of $20$ full timeslots per transfer. For heavy-tailed distribution that is based on Pareto distribution, we used a minimum transfer size equal to that of $2$ full timeslots. Finally, to prevent generation of intractably large transfers, we limited maximum transfer volume to that of $2000$ full timeslots. We focus on scenarios with no link failures to evaluate gains.

\begin{table}
\caption{Various topologies used in evaluation} \label{table_topology}
\begin{center}
\begin{tabular}{ |p{1.5cm}|p{6.3cm}| }
    \hline
    \textbf{Name} & \textbf{Description} \\
    \hline
    \hline
    Random & Randomly generated and strongly connected with $50$ nodes and $150$ edges. Each node has a minimum connectivity of two. \\
    \hline
    GScale \cite{b4} & Connects Google datacenters across the globe with $12$ nodes and $19$ links. \\
    \hline
    Cogent \cite{cogent} & A large backbone and transit network that spans across USA and Europe with $197$ nodes and $243$ links. \\
    \hline
\end{tabular}
\end{center}
\end{table}

\begin{table}
\caption{Transfer size distributions used in evaluation} \label{table_traffic}
\begin{center}
\begin{tabular}{ |p{2cm}|p{5.8cm}| }
    \hline
    \textbf{Name} & \textbf{Description} \\
    \hline
    \hline
    Light-tailed & According to Exponential distribution. \\
    \hline
    Heavy-tailed & According to Pareto distribution. \\
    \hline
    Facebook \textit{Cache-Follower} \cite{social_inside} & Generated across Facebook inter-datacenter networks running cache applications. \\
    \hline
    Facebook \textit{Hadoop} \cite{social_inside} & Generated across Facebook inter-datacenter networks running geo-distributed analytics. \\
    \hline
\end{tabular}
\end{center}
\end{table}

\subsection{Comparison of Scheduling Policies over Forwarding Trees}
\begin{figure}[t!]
\centering
\includegraphics[width=\columnwidth]{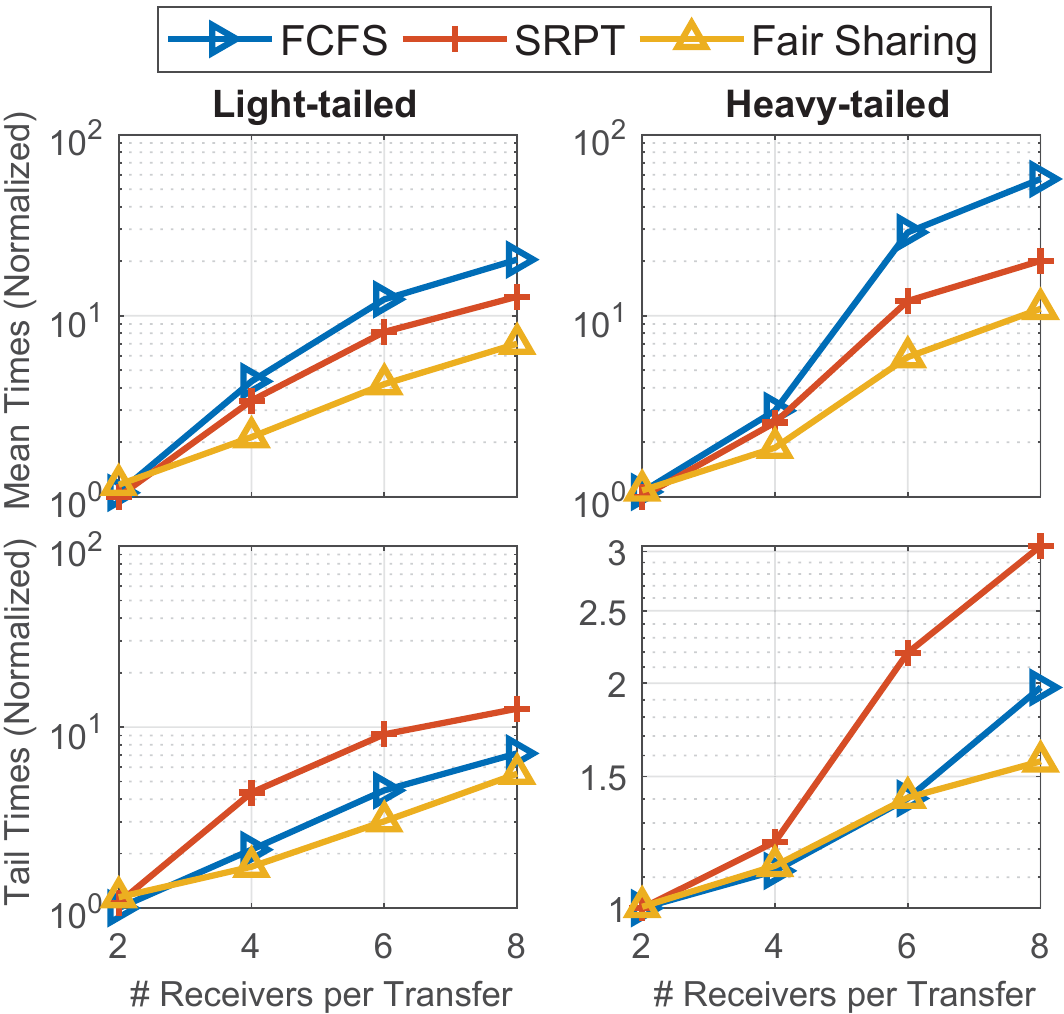}
\caption{Performance of three well-known scheduling policies under heavy load (forwarding tree selection according to DCCast)}
\label{fig_policies}
\end{figure}

We first compare the performance of three well-known scheduling policies of FCFS, SRPT and Fair Sharing (based on MMF). We used the weight assignment in \cite{dccast} for forwarding tree selection and considered Random topology in Table \ref{table_topology}. We considered both light-tailed and heavy-tailed distributions. All policies used almost identical amount of bandwidth (not shown). Under light load, we obtained results similar to scheduling traffic on a single link where SRPT performs better than Fair Sharing (not shown). Figure \ref{fig_policies} shows the results of our experiment under heavy load. When the number of receivers is small, SRPT is the best policy to minimize mean times. However, as we increase the number of receivers (larger trees), Fair Sharing offers better mean and tail times. This simply occurs because the contention due to overlapping trees caused by prioritizing transfers over one another (either according to residual size in case of SRPT or arrival order in case of FCFS) increases as more transfers are submitted or as transfers grow in size.

\subsection{Bandwidth usage of partitioning techniques}
\begin{figure*}[t!]
\centering
\includegraphics[width=\textwidth]{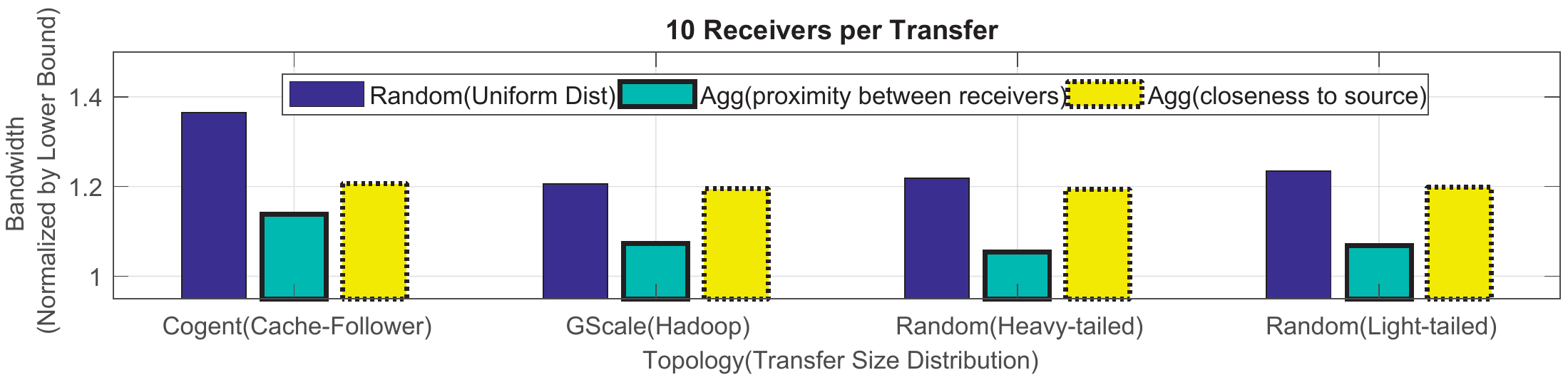}
\caption{Bandwidth usage of various partitioning techniques (lower is better)}
\label{fig_bw}
\end{figure*}

We considered three partitioning techniques and measured the average bandwidth usage over multiple runs and many timeslots. We used the topologies in Table \ref{table_topology} and traffic patterns of Table \ref{table_traffic}. Figure \ref{fig_bw} shows the results. We calculated the lower bound by considering a single minimal edge Steiner tree per transfer. Other schemes considered are: \textit{Random(Uniform Dist)} breaks each set of receivers into two partitions by randomly assigning each receiver to one of the two partitions with equal probability, \textit{Agg(proximity between receivers)} clusters receivers according to closeness to each other, and \textit{Agg(closeness to source)} clusters receivers according to their distance from source (receivers closer to source are bundled together). As can be seen, \textit{Agg(proximity between receivers)}, which is used by QuickCast, provides the least bandwidth overhead (up to $17\%$ to lower bound). In general, breaking receivers into subsets that are attached to a sender with minimal bandwidth usage is an open problem.

\subsection{QuickCast with different partitioning approaches}
\begin{figure*}
\centering
\includegraphics[width=\textwidth]{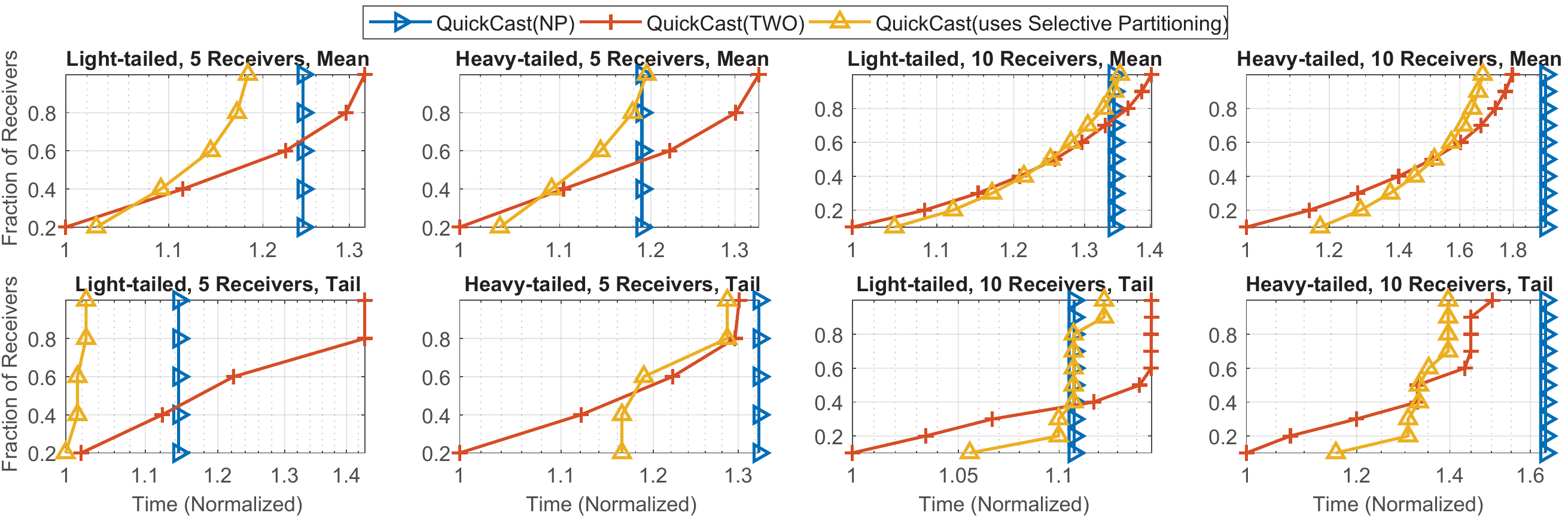}
\caption{Comparison of partitioning approaches in Table \ref{table_schemes}}
\label{fig_dca}
\end{figure*}

We compare three partitioning approaches shown in Table \ref{table_schemes}. We considered receiver sets of $5$ and $10$ with both light-tailed and heavy-tailed distributions. We show both mean (top row) and tail (bottom row) completion times in the form of a CDF in Figure \ref{fig_dca}. As expected, when there is no partitioning, all receivers complete at the same time (vertical line in CDF). When partitioning is always applied, completion times can jump far beyond the no partitioning case due to unnecessary creation of additional weak links. The benefit of QuickCast is that it applies partitioning selectively. The amount of benefit obtained is a function of partitioning factor $p_f$ (which for the topologies and traffic patterns considered here was found to be most effective between $1.0$ and $1.1$ according to our experiments, we used $1.1$). With QuickCast, the fastest receiver can complete up to $41\%$ faster than the slowest receiver and even the slowest receiver completes up to $10\%$ faster than when no partitioning is applied.

\begin{table}
\caption{Evaluation of partitioning techniques for P2MP transfers} \label{table_schemes}
\begin{center}
\begin{tabular}{ |p{2cm}|p{5.5cm}| }
    \hline
    \textbf{Scheme} & \textbf{Details} \\
    \hline
    \hline
    QuickCast & Algorithm \ref{algo_quick} (\textit{Selective Partitioning}). \\
    \hline
    QuickCast(NP) & QuickCast with no partitioning applied. \\
    \hline
    QuickCast(TWO) & QuickCast with two partitions always. \\
    \hline
\end{tabular}
\end{center}
\end{table}

\subsection{QuickCast vs. DCCast}
\begin{figure}[t!]
\centering
\subfigure[Heavy Load]{
\includegraphics[width=\columnwidth]{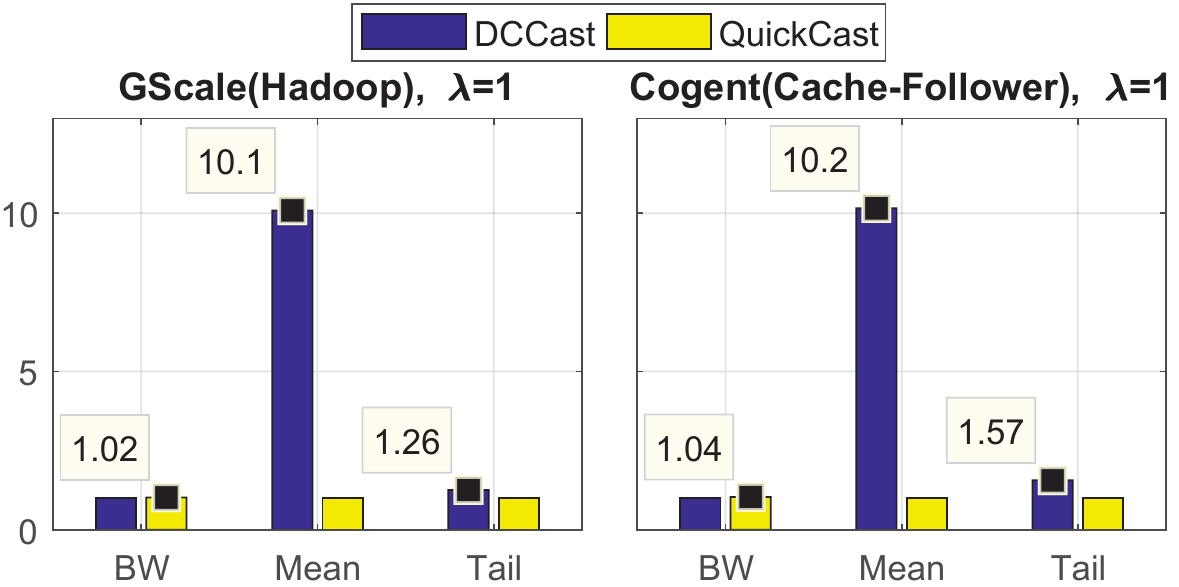}
}

\subfigure[Light Load]{
\includegraphics[width=\columnwidth]{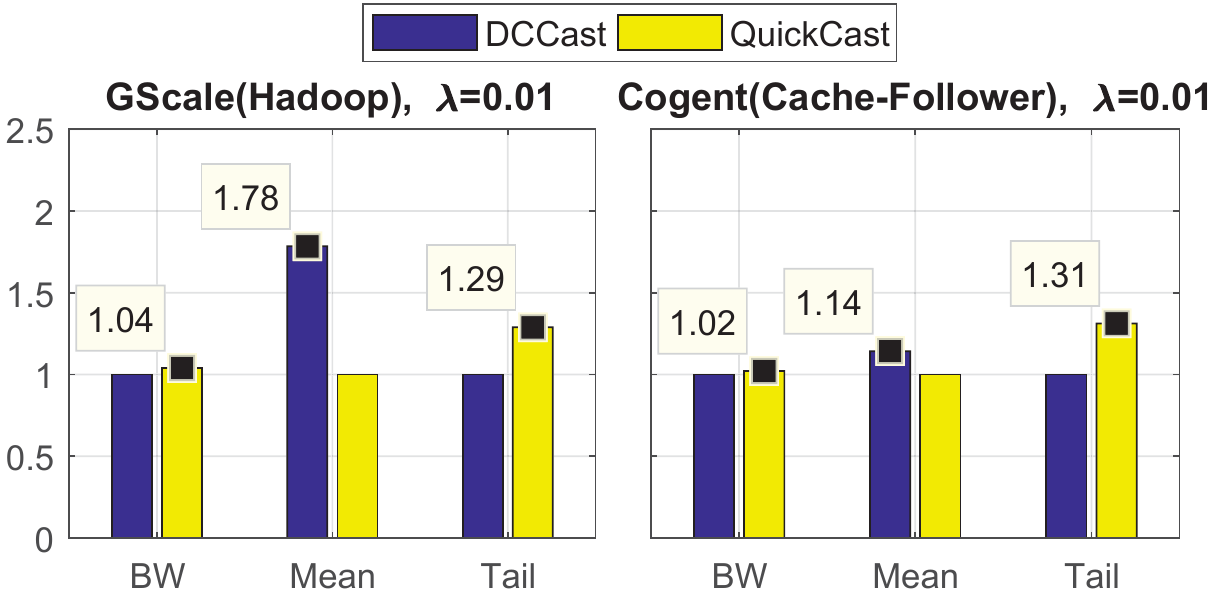}
}
\caption{Comparison of completion times of and bandwidth used by QuickCast vs. DCCast (Normalized by minimum in each category)} \label{fig_s1}
\end{figure}

We now compare QuickCast with DCCast using real topologies and inter-datacenter transfer size distributions, namely \textbf{GScale(Hadoop)} and \textbf{Cogent(Cache-Follower)} shown in Tables \ref{table_topology} and \ref{table_traffic}. Figure \ref{fig_s1} shows the results. We considered $10$ receivers per P2MP transfer. In all cases, QuickCast uses up to $4\%$ more bandwidth. For lightly loaded scenarios where $\lambda = 0.01$, QuickCast performs up to $78\%$ better in mean times, but about $35\%$ worse in tail times. The loss in tail times is a result of rate-allocation policy: FCFS performs better in tail times compared to Fair Sharing under light loads where contention due to overlapping trees is negligible (similar to single link case when all transfers compete for one resource). For heavily loaded scenarios where $\lambda = 1$, network contention due to overlapping trees is considerable and therefore QuickCast has been able to reduce mean times by about $10\times$ and tail times by about $57\%$. This performance gap continues to increase in favor of QuickCast as offered load grows further. In general, operators aim to maximize network utilization over dedicated WANs \cite{b4, swan} which could lead to heavily loaded time periods. Such scenarios may also appear as a result of bursty transfer arrivals.

\begin{figure}[t!]
\centering
\subfigure[Cogent(Cache-Follower)]{
\includegraphics[width=\columnwidth]{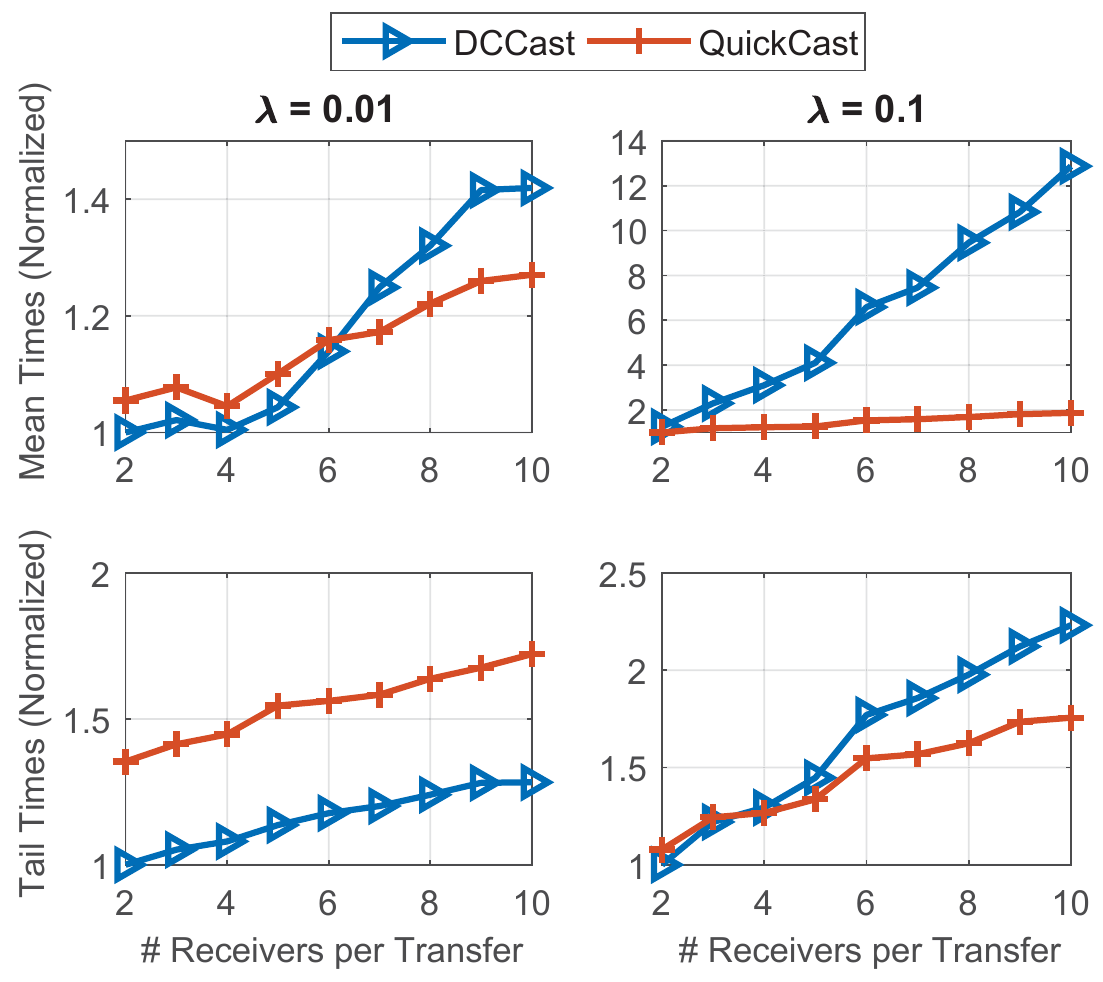}
}

\subfigure[GScale(Hadoop)]{
\includegraphics[width=0.98\columnwidth]{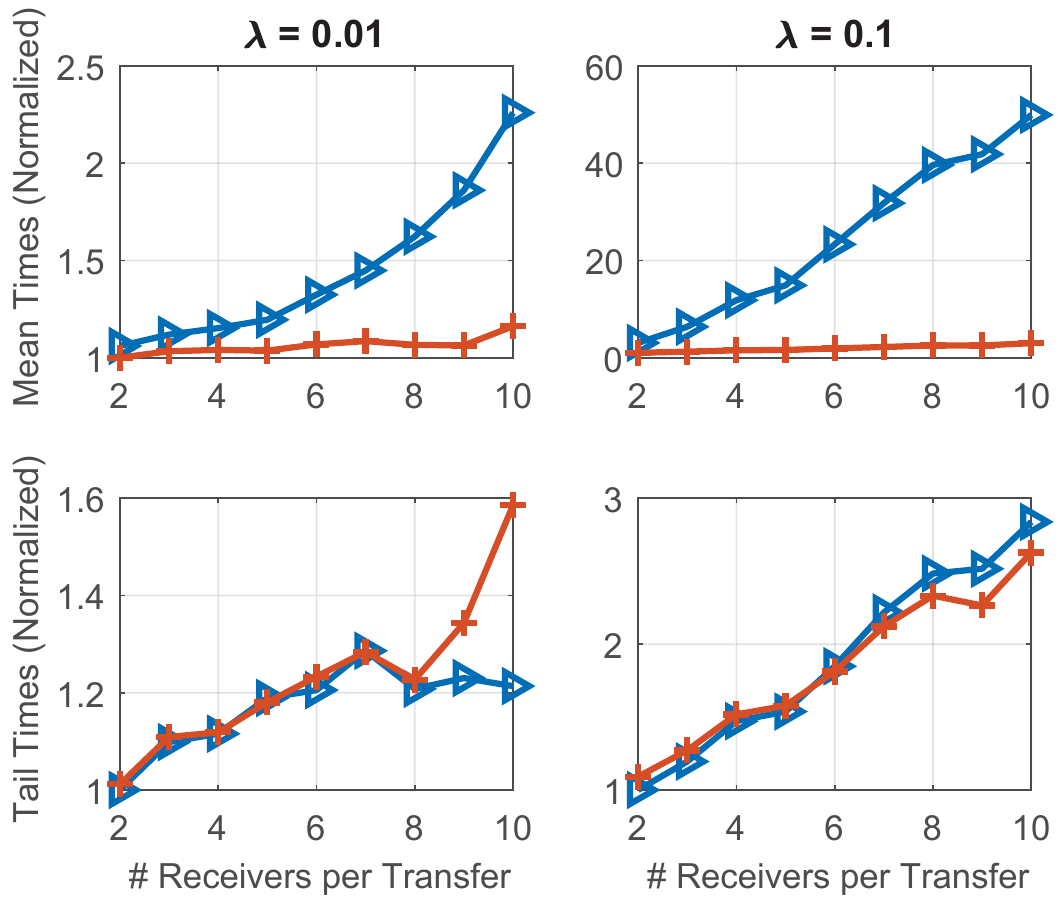}
}
\caption{Comparison of completion times of QuickCast and DCCast (Normalized by minimum in each category) by number of object copies} \label{fig_s2}
\end{figure}

In a different experiment, we studied the effect of number of replicas of data objects on performance of DCCast and QuickCast as shown in Figure \ref{fig_s2} (please notice the difference in vertical scale of different charts). Bandwidth usage of both schemes were almost identical (QuickCast used less than $4\%$ extra bandwidth in the worst case). We considered two operating modes of lightly to moderately loaded ($\lambda = 0.01$) and moderately to heavily loaded ($\lambda = 0.1$). QuickCast offers most benefit when number of copies grows. When the number of copies is small, breaking receivers into multiple sets may provide limited benefit or even degrade performance as resulting partitions will be too small. This is why mean and tail times degrade by up to $5\%$ and $40\%$ across both topologies and traffic patterns when network is lightly loaded, respectively. Under increasing load and with more copies, it can be seen that QuickCast can reduce mean times significantly, i.e., by as much as $6\times$ for \textbf{Cogent(Cache-Follower)} and as much as $16\times$ for \textbf{GScale(Hadoop)}, respectively. The sudden increase in tail times for GScale topology is because this network has only $12$ nodes which means partitioning while making $10$ copies may most likely lead to overlapping edges across partitioned trees and increase completion times. To address this problem, one could reduce $p_f$ to minimize unnecessary partitioning.

\subsection{Complexity}
We discuss two complexity metrics of run-time and number of Group Table entries needed to realize QuickCast.

\textbf{Computational Complexity:} We computed run-time on a machine with a Core-i7 6700 CPU and 24GBs of memory using JRE 8. We used the same transfer size properties mentioned at the beginning of this section. With \textbf{GScale(Hadoop)} setting and $\lambda = 0.01$, run-time of procedure \textbf{Submit} increased from $1.44ms$ to $2.37ms$ on average while increasing copies from $2$ to $10$. With the same settings, runtime of procedure \textbf{DispatchRates} stayed below $2\mu s$ for varying number of copies. Next, we increased both load and network size by switching to \textbf{Cogent(Cache-Follower)} setting and $\lambda = 1.0$. This time, run-time of procedure \textbf{Submit} increased from $3.5ms$ to $35ms$ and procedure \textbf{DispatchRates} increased from $0.75ms$ to $1.6ms$ on average while increasing copies from $2$ to $10$. Although these run-times are significantly smaller than timeslot widths used in prior works which are in the range of minutes \cite{amoeba, owan}, more efficient implementation of proposed techniques may result in even further reduction of run-time. Finally, with our implementation, memory usage of QuickCast algorithm is in the order of $100$'s of megabytes on average.

\textbf{Group Table Entries:} We performed simulations over $2000$ timeslots with $\lambda = 1.0$ (heavily loaded scenario) and number of copies set to $10$. With \textbf{GScale(Hadoop)} setting, the maximum number of required Group Table entries was $455$ and the average of maximum rules for all nodes observed over all timeslots was $166$. With \textbf{Cogent(Cache-Follower)} setting, which is more than $10$ times larger than GScale, we observed a maximum of $165$ and an average of $9$ Group Table entries for the maximum observed by all nodes over all timeslots. We considered the highly loaded case as it leads to higher number of concurrent forwarding trees. Currently, most switches that support Group Tables offer a maximum of $512$ or $1024$ entries in total. In this experiment, a maximum of one Group Table entry per transfer per node was enough as we considered switches that support up to $32$ action buckets per entry \cite{of-juniper-explain} which is more then total number of receivers we chose per transfer. In general, we may need more than one entry per node per transfer or we may have to limit the branching factor of selected forwarding trees, for example when Group Table entries support up to $8$ action buckets \cite{of-hp-3}.

\section{Discussion} \label{discussion}
The focus of this paper is on algorithm design and abstract evaluations. In this section, we dive a bit further into practical details and issues that may arise in a real setting.

\textbf{Timeslot duration:} One configuration factor is timeslot length $\delta$. In general, smaller timeslots allow for faster response to changes and arrival of new requests, but add the overhead of rate computations. Minimum possible timeslot length depends on how long it takes for senders to converge to centrally allocated rates.

\textbf{Handling rate-limiting inaccuracies:} Rate-limiting is generally not very accurate, especially if done in software \cite{dctc}. To deal with inaccuracies and errors, every sender has to report back to the TE server at the end of every timeslot and specify how much traffic it was able to deliver. TE server will deduct these from the residual demand of requests to get the new residual demands. Rate-allocation continues until a request is completely satisfied.

\textbf{Receiver feedback to sender:} Forwarding trees allow flow of data from sender to receivers but receivers also need to communicate with senders. Forwarding rules can be installed so that it supports receivers sending feedback back to sender over the same tree but in reverse direction. There will not be need to use Group Tables on the reverse direction since no replication is performed. One can use a simple point to point scheme for receiver feedback. Since we propose applying forwarding trees over wired networks with rate-limiting, dropped packets due to congestion and corruptions are expected to be low. This means if a scheme such as Negative Acknowledgement (NAK) is used, receiver feedback should be tiny and can be easily handled by leaving small spare capacity over edges.

\textbf{Handling network capacity loss:} Link/switch failures may occur in a real setting. In case of a failure, the TE server can be notified by a network element that detects the failure. The TE server can then exclude the faulty link/switch from topology and re-allocate all requests routed on that link using their residual demands. After new rates are allocated and forwarding trees recomputed, forwarding plane can be updated and new rates can be given to end-points for rate-limiting.

\textbf{TE server failure:} Another failure scenario is when TE server stops working. It is helpful if end-points are equipped with some distributed congestion control mechanism. In case TE server fails, end-points can roll back to the distributed mode and determine their rates according to network feedback.

\section{Related Work} \label{related_works}
IP multicasting \cite{ip_multicast}, CM \cite{centralized-multicast}, TCP-SMO \cite{tcp-smo} and NORM \cite{norm} are instances of approaches where receivers can join groups anytime to receive required data and multicast trees are updated as nodes join or leave. This may lead to trees far from optimal. Also, since load distribution is not taken into account, network capacity may be poorly utilized. 

Having knowledge of the topology, centralized management allows for more careful selection of multicast trees and improved utilization via rate-control and bandwidth reservation. CastFlow \cite{castflow} precalculates multicast trees which can then be used at request arrival time for rule installation. ODPA \cite{odpa} presents algorithms for dynamic adjustment of multicast spanning trees according to specific metrics. These approaches however do not apply rate-control. MPMC \cite{MPMC_2013, MPMC_2016} proposes use of multiple multicast trees for faster delivery of files to many receivers then applies coding for improved performance. MPMC does not consider the inter-play between transfers when many P2MP requests are initiated from different source datacenters. In addition, MPMC requires continues changes to the multicast tree which incurs significant control plane overhead as number of chunks and transfers increases. \cite{sdn-based-file-dist-rate} focuses on rate and buffer size calculation for senders. This work does not propose any solution for tree calculations. RAERA \cite{raera} is an approximation algorithm to find Steiner trees that minimize data recovery costs for multicasting given a set of recovery nodes. We do not have recovery nodes in our scenario. MTRSA \cite{sdn_multicast} is an approximation algorithm for multicast trees that satisfy a minimum available rate over a general network given available bandwidth over all edges. This work assumes constant rate requirement for transfers and focuses on minimizing bandwidth usage rather than completion times. 

For some regular and structured topologies, such as FatTree intra-datacenter networks, it is possible to find optimal (or close to optimal) multicast trees efficiently. Datacast \cite{datacast} sends data over edge-disjoint Steiner trees found by pruning spanning trees over various topologies of FatTree, BCube and Torus. AvRA \cite{avalanche} focuses on Tree and FatTree topologies and builds minimal edge Steiner trees that connect the sender to all receivers as they join. MCTCP \cite{mctcp} reactively schedules flows according to link utilization.

As an alternative to in-network multicasting, one can use overlay networks where hosts perform forwarding. RDCM \cite{rdcm} populates backup overlay networks as nodes join and transmits lost packets in a peer-to-peer fashion over them. NICE \cite{nice} creates hierarchical clusters of multicast peers and aims to minimize control traffic overhead. AMMO \cite{AMMO} allows applications to specify performance constraints for selection of multi-metric overlay trees. DC2 \cite{dc2} is a hierarchy-aware group communication technique to minimize cross-hierarchy communication. SplitStream \cite{split-stream} builds forests of multicast trees to distribute load across many machines. Due to lack of complete knowledge of underlying network topology and status (e.g. link utilizations, congestion or even failures), overlay systems are limited in reducing bandwidth usage and managing distribution of traffic.

Alternatives to multicasting for bulk data distribution include peer-to-peer \cite{promise, bittorrent, slurpie} and store-and-forward \cite{netstitcher, mbdt, dtb, mbdt_initial} approaches. Peer-to-peer approaches do not consider careful link level traffic allocation and scheduling and do not focus on minimizing bandwidth usage. The main focus of peer-to-peer schemes is to locally and greedily optimize completion times rather than global optimization over many transfers across a network. Store-and-forward approaches focus on minimizing costs by utilizing diurnal traffic patterns while delivering bulk objects and incur additional bandwidth and storage costs on intermediate datacenters. Coflows are another related concept where flows with a collective objective are jointly scheduled for improved performance \cite{coflow}. Coflows however do not aim at bandwidth savings.

There are recent solutions for management of P2MP transfers with deadlines. DDCCast \cite{ddccast} uses a single forwarding tree and the As Late As Possible (ALAP) scheduling policy for admission control of multicast deadline traffic. In \cite{ji_siqi}, authors propose use of few parallel forwarding trees from source to all receivers to increase throughput and meet more deadlines considering transfer priorities and volumes. The techniques we proposed in QuickCast for receiver set partitioning can be applied to these prior work for further performance gains.

In design of QuickCast, we did not focus on throughput-optimality, which guarantees network stability for any offered load in capacity region. We believe our tree selection approach, which balances load across many existing forwarding trees, aids in moving towards throughput-optimality. One could consider developing a P2MP joint scheduling and routing scheme with throughput-optimality as the objective. However, in general, throughput-optimality does not necessarily lead to highest performance (e.g., lowest latency) \cite{yekkehkhany2017near, yekkehkhany2017gb}.

\textbf{Reliability:} Various techniques have been proposed to make multicasting reliable including use of coding and receiver (negative or positive) acknowledgements or a combination of them \cite{reliable-multicasting}. In-network caching has also been used to reduce recovery delay, network bandwidth usage and to address the ACK/NAK implosion problem \cite{datacast, arm}. Using positive ACKs does not lead to ACK implosion for medium scale (sub-thousand) receiver groups \cite{tcp-smo}. TCP-XM \cite{tcp-xm} allows reliable delivery by using a combination of IP multicast and unicast for data delivery and re-transmissions. MCTCP \cite{mctcp} applies standard TCP mechanisms for reliability. Receivers may also send NAKs upon expiration of some inactivity timer \cite{norm}. NAK suppression to address implosion can be done by routers \cite{arm}. Forward Error Correction (FEC) can be used for reliable delivery, using which a sender encodes $k$ pieces of an object into $n$ pieces ($k \le n$) where any $k$ out of $n$ pieces allow for recovery \cite{fec, alc}. FEC has been used to reduce re-transmissions \cite{norm} and improve the completion times \cite{avalanche_code}. Some popular FEC codes are Raptor Codes \cite{raptor} and Tornado Codes \cite{tornado}.

\textbf{Congestion Control:} PGMCC \cite{pgmcc}, MCTCP \cite{mctcp} and TCP-SMO \cite{tcp-smo} use window-based TCP like congestion control to compete fairly with other flows. NORM \cite{norm} uses an equation-based rate control scheme. Datacast \cite{datacast} determines the rate according to duplicate interest requests for data packets. All of these approaches track the slowest receiver.

\section{Conclusions and Future Work}
In this paper, we presented QuickCast algorithm to reduce completion times of P2MP transfers across datacenters. We showed that by breaking receiver sets of P2MP transfers with many receivers into smaller subsets and using a separate tree per subset, we can reduce completion times. We proposed partitioning according to proximity as an effective approach for finding such receiver subsets, and showed that partitioning need be applied to transfers selectively. To do so, we proposed a partitioning factor that can be tuned according to topology and traffic distribution. Further investigation is required on finding metrics to selectively apply partitioning per transfer. Also, investigation of partitioning techniques that optimize network performance metrics as well as study of optimality bounds of such techniques are left as part of future work. 

Next, we discovered that while managing P2MP transfers with many receivers, FCFS policy used in DCCast creates many contentions as a result of overlapping forwarding trees significantly reducing utilization. We applied Fair Sharing policy reducing network contention and improving completion times. Finally, we performed experiments with well-known rate-allocation policies and realized that Max-Min Fairness provides much lower completion times compared to SRPT while scheduling over large forwarding trees. More research is needed on best rate-allocation policy for P2MP transfers. Alternatively, one may drive a more effective joint partitioning, rate-allocation and forwarding tree selection algorithm by approximating a solution to optimization model we proposed.

\bibliographystyle{IEEEtran}
\bibliography{citations}

\end{document}